\documentstyle[aps,epsfig]{revtex}
\newcommand{\be}{\begin{displaymath}}
\newcommand{\ee}{\end{displaymath}}
\begin{document}
\hyphenation{wave-number}

\title{Bogoliubov dispersion relation and the possibility of 
superfluidity for weakly-interacting photons in a 2D photon fluid} 
\author{Raymond Y. Chiao and Jack Boyce\\
Dept. of Physics, University of California, Berkeley, 
California 94720-7300\\
E-mail addresses: 
chiao@physics.berkeley.edu, jboyce@physics.berkeley.edu}
\date{Received 3 May, 1999; revised manuscript received 11 July, 
1999; galley proof version of October 11, 1999}
\maketitle
\begin{abstract}
The Bogoliubov dispersion relation for the elementary excitations of 
the weakly-interacting Bose gas is shown to hold for the case of the 
weakly-interacting photon gas (the ``photon fluid'') in a nonlinear 
Fabry-Perot cavity.  The chemical potential of a photon in the 2D 
photon fluid does not vanish.  The Bogoliubov relation, which is also 
derived by means of a linearized fluctuation analysis in classical 
nonlinear optics, implies the possibility of a new, superfluid state 
of light.  The theory underlying an experiment in progress to observe 
sound waves in the photon fluid is described, and another experiment to 
measure the critical velocity of this superfluid 
is proposed.\\
\end{abstract}
\section{Introduction}

 The quantum many-body problem, with its many, rich manifestations in 
 condensed matter physics, has had a long and illustrious history.  In 
 particular, superconductivity and superfluidity were two major 
 discoveries in this field.  Although at present much is well 
 understood (e.g., the BCS theory of superconductivity), the recent 
 experimental discoveries of Bose-Einstein condensation in 
 laser-cooled atoms~\cite{Cornell,Hulet,Ketterle,Walls} raises new and 
 interesting questions, such as whether the observed Bose-Einstein 
 condensates are superfluids, or whether persistent currents can exist 
 in these new states of matter.
 
Historically speaking, in the study of the interaction of light with 
matter, most of the emphasis has been on exploring new states of 
matter, such as the recently observed atomic Bose-Einstein condensates.  
However, not as much attention has been focused on exploring new 
states of light.  Of course, the invention of the laser led to the 
discovery of a new state of light, namely the coherent state, which is 
a very robust one.  Two decades ago, squeezed states were discovered, 
but these states are not as robust as the coherent state, since they 
are easily degraded by scattering and absorption.  In contrast to the 
laser, which involves a system far away from equilibrium, we shall 
explore here states close to the ground state of a photonic system.  
Hence they should be robust ones.
 
 Here we shall study the many-body problem by studying the interacting 
 many-photon system (the ``photon fluid'') near its ground state.  In 
 this paper we shall explore some theoretical considerations which 
 suggest the possibility of a new state of light, namely, the 
 superfluid state.  In particular, we shall derive the Bogoliubov 
 dispersion relation for the weakly-interacting photon gas with 
 repulsive photon-photon interactions, starting both from the 
 microscopic (i.e., second-quantized) level, and also from the 
 macroscopic (i.e., classical-field) level.  Thereby we shall find an 
 expression for the effective chemical potential of a photon in the 
 photon fluid, 
 and shall relate the velocity of sound in the photon fluid to this 
 nonvanishing chemical potential.
 In this way, we lay the theoretical foundations for an 
 experiment in progress to measure the sound-wave-like dispersion 
 relation for the photon fluid.  We also propose another experiment to
measure the critical velocity of this fluid, and thus to 
 test for the possibility of the superfluidity of the resulting state of 
 the light.
 
 Although the interaction Hamiltonian used in this paper is equivalent 
 to that used earlier in four-wave squeezing, we emphasize here the 
 many-body, collective aspects of the problem which result from 
 multiple photon-photon interactions.  This leads to the idea of the 
 ``photon fluid.''  Since the microscopic and macroscopic analyses 
 yield the same Bogoliubov dispersion relation for excitations of this 
 fluid, it may be argued that there is nothing fundamentally new in the 
 microscopic analysis given below which is not already contained in 
 the macroscopic, classical nonlinear optical analysis.  However, it 
 is the microscopic analysis which leads to the new, heuristic 
 viewpoint of the interacting photon system as a ``photon fluid,'' a 
 conception which could give rise to new ways of understanding and 
 discovering nonlinear optical phenomena.  Furthermore, the 
 interesting question of the 
 quantum optical state of the light inside
 the cavity resulting from multiple interactions between the photons 
 (i.e., whether it results in a coherent, squeezed, Fock, or some 
 other quantum state), cannot be addressed by classical nonlinear 
 optical methods.  Thus this paper represents a first attempt to 
 formulate the new concept of a ``photon fluid'' starting from the 
 microscopic viewpoint, and to lay the foundations for answering the 
 question concerning 
 the resulting quantum optical state of the
 light.  

\section{The Bogoliubov problem}

Here we re-examine one particular many-body problem, the one first 
solved by Bogoliubov~\cite{Bogoliubov=1947,Pines}.  Suppose that one 
has a zero-temperature system of bosons which are interacting with each other 
repulsively, for example, a dilute system of small, bosonic hard 
spheres.  Such a model was intended to describe superfluid helium, but 
in fact it did not work well there, since the interactions between 
atoms in superfluid helium were too strong for the theory for be 
valid.  In order to make the problem tractable theoretically, let us 
assume that these interactions are weak.  In the case of light, the 
interactions between the photons are in fact always weak, so that this 
assumption is a good one.  However, these interactions are
nonvanishing, as demonstrated by the fact that photon-photon collisions
mediated by atoms excited near, but off, resonance have been experimentally 
observed~\cite{Morgan}.  We start with the Bogoliubov Hamiltonian
\begin{eqnarray}
H &=& H_{free}+H_{int}\nonumber\\
H_{free}&=&\sum_{p} \epsilon(p)a_{p}^{\dagger}a_{p}\nonumber\\
H_{int}&=&\frac{1}{2}\sum_{\kappa  pq} 
 V(\kappa  ) a_{p+\kappa  }^{\dagger} a_{q-\kappa  }^{\dagger}a_{p}a_{q}\,,
 \label{1}
\end{eqnarray}
where the operators $a_{p}^{\dagger}$ and $a_{p}$ are 
creation and annihilation operators, respectively, for bosons with 
momentum $p$, which satisfy the Bose commutation relations
\begin{equation}
[a_{p},a_{q}^{\dagger}] = \delta_{pq}\mbox{ and } [a_{p},a_{q}] = 
[a_{p}^{\dagger},a_{q}^{\dagger}] = 0\,.
\end{equation}

The first term $H_{free}$ in the Hamiltonian represents the energy of 
the free boson system, and the second term $H_{int}$ represents the 
energy of the interactions between the bosons arising from the 
potential energy $V(\kappa  )$.  The interaction term is 
equivalent to the one 
responsible 
for producing squeezed states of light via four-wave 
mixing~\cite{Slusher}.  It represents the annihilation of two 
particles, here photons, of momenta $p$ and $q$, along with the 
creation of two particles with momenta $p+\kappa  $ and $q-\kappa  $, in other 
words, a scattering process with a momentum transfer $\kappa  $ between a 
pair of particles with initial momenta $p$ and $q$, along with the 
assignment of an energy $V(\kappa  )$ to this scattering process. 

\section{The free-photon dispersion relation inside a Fabry-Perot 
resonator}

\begin{figure}
\centerline{\psfig{figure=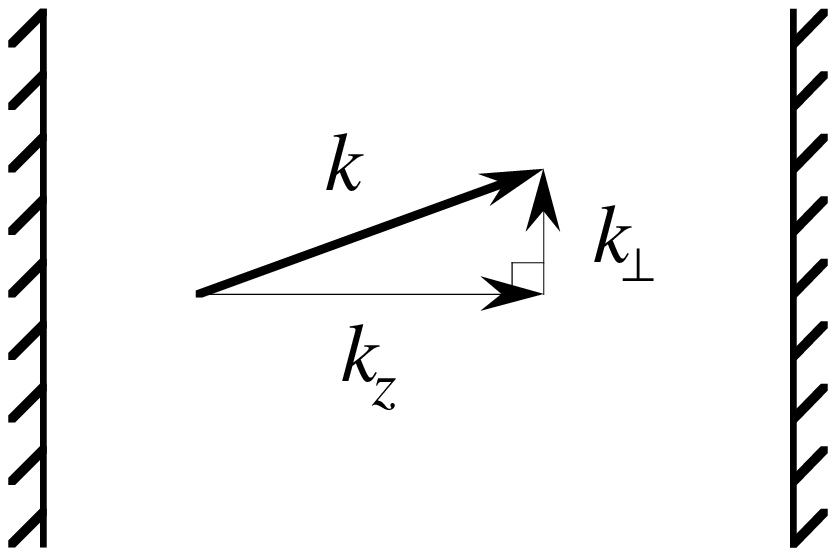,width=7cm}}
\caption{A planar Fabry-Perot imposes boundary conditions which 
quantize the allowed values of $k_{z}$, where $z$ is the axis normal 
to the mirrors, in units of $\pi/L$, where $L$ is the separation of 
the mirrors.  For a plane-wave mode which propagates at a small angle 
with respect to the $z$ axis, there arises an effective 
nonrelativistic energy-momentum relation for an noninteracting, 
trapped 2D photon, whose effective mass is $m=\hbar\omega/c^{2}$ (see 
text).}
\label{Fabry-Perot}
\end{figure}

Photons with momenta $p$ and $q$ also obey the above commutations 
relations, so that the Bogoliubov theory should in principle also 
apply to the weakly-interacting photon gas.  The factor 
$\epsilon(p)$ represents the energy as a function of the momentum 
(the dispersion relation) for the free, i.e., noninteracting, bosons.  In the 
case of photons in a Fabry-Perot resonator, the boundary conditions of 
the mirrors cause the $\epsilon(p)$ of a photon trapped inside the 
resonator to correspond to an energy-momentum relation which is 
identical to that of a nonrelativistic particle with an effective 
mass~\cite{Morgan,Garrison} of $m=\hbar\omega/c^{2}$.  This can be 
understood starting from Fig.~\ref{Fabry-Perot}.
 
For high-reflectivity mirrors, the vanishing of the electric field at 
the reflecting surfaces of the mirrors imposes a quantization 
condition on the allowed values of the $z$-component of the photon 
wave vector, $k_{z}=n\pi/L$, where $n$ is an integer, and $L$ is the 
distance between the mirrors.  Thus the usual frequency-wavevector 
relation
\begin{equation}
\omega(k) = c[k_{x}^{2}+k_{y}^{2}+k_{z}^{2}]^{1/2}\,,
\end{equation}
upon multiplication by $\hbar$, becomes the energy-momentum 
relation for the photon 
\begin{equation}
E(p) = c[p_{x}^{2}+p_{y}^{2}+p_{z}^{2}]^{1/2} = 
c[p_{x}^{2}+p_{y}^{2}+\hbar^{2}n^{2}\pi^{2}/L^{2}]^{1/2}
=c[p_{x}^{2}+p_{y}^{2}+m^{2}c^{2}]^{1/2}\,,
\end{equation}
where $m={\hbar}n\pi/Lc$ is the effective mass of the 
photon.  In the limit of small-angle (or paraxial) propagation, where 
the small transverse momentum of the photon satisfies the inequality
\begin{equation}
p_{\perp}= [p_{x}^{2}+p_{y}^{2}]^{1/2}\ll p_{z}=\hbar k_{z} = \hbar 
n\pi/L\,,
\end{equation}
we obtain from a Taylor expansion of the relativistic relation, a 
nonrelativistic energy-momentum relation for the 2D noninteracting 
photons inside the Fabry-Perot resonator
\begin{equation}
E(p_{\perp}) \cong mc^{2}+p_{\perp}^{2}/2m\,,
\label{nonrelativistic}
\end{equation}
where $m={\hbar}n\pi/Lc\cong \hbar\omega/c^{2}$ is the effective mass 
of the confined photons.  It is convenient to redefine the zero of 
energy, so that only the effective kinetic energy,
\begin{equation}
\epsilon(p_{\perp}) \cong p_{\perp}^{2}/2m\,,
\label{7}
\end{equation}
remains.  To establish the connection with the Bogoliubov Hamiltonian, 
we identify the two-dimensional momentum $p_{\perp}$ as the momentum 
$p$ that appears in this Hamiltonian, and the above $\epsilon(p_{\perp})$ 
as the $\epsilon(p)$ that appears in Eq.~(\ref{1}).

\section{The Bogoliubov dispersion relation for the photon fluid}

Now we know that in an ideal Bose gas at absolute zero temperature, 
there exists a Bose condensate consisting of a macroscopic number 
$N_{0}$ of particles occupying the zero-momentum state.  This feature 
should survive in the case of the weakly-interacting Bose gas, since 
as the interaction vanishes, one should recover the Bose condensate 
state.  Hence following Bogoliubov, we shall assume here that even in 
the presence of interactions, $N_{0}$ will remain a macroscopic number 
in the photon fluid\cite{Thouless}.
This macroscopic number will be determined by the 
intensity of the incident laser beam which excites the Fabry-Perot 
cavity system, and turns out to be a very large number compared to 
unity (see below).  For the ground state wave function 
$\Psi_{0}(N_{0})$ with $N_{0}$ particles in the Bose condensate in the 
$p=0$ state, the zero-momentum operators $a_{0}$ and $a_{0}^{\dagger}$ 
operating on the ground state obey the relations
\begin{eqnarray}
a_{0}\left|\Psi_{0}(N_{0})\right>&=&
\sqrt{N_{0}}\left|\Psi_{0}(N_{0}-1)\right>\nonumber \\
a_{0}^{\dagger}\left|\Psi_{0}(N_{0})\right>&=&
\sqrt{N_{0}+1}\left|\Psi_{0}(N_{0}+1)\right>\,.
\end{eqnarray}
Since $N_{0}\gg1$, we shall neglect the difference between the factors 
$\sqrt{N_{0}+1}$ and $\sqrt{N_{0}}$.  
Thus one can 
replace all occurrences of $a_{0}$ and $a_{0}^{\dagger}$ by the 
$c$-number $\sqrt{N_{0}}$, so that to a good approximation 
$[a_{0},a_{0}^{\dagger}]\approx 0$.  
However, the number of particles in 
the system is then no longer exactly conserved, as can be seen by 
examination of the term in the Hamiltonian
\begin{equation}
\sum_{\kappa }V(\kappa )a_{\kappa }^{\dagger}a_{-\kappa 
}^{\dagger}a_{0}a_{0}\approx N_{0}\sum_{\kappa }V(\kappa )a_{\kappa 
}^{\dagger}a_{-\kappa }^{\dagger}\,,
\end{equation}
which represents the creation of a pair of particles, i.e., photons, with 
momenta $\kappa  $ and $-\kappa  $ out of nothing.  

However, whenever the system is open one, i.e., whenever it is 
connected to an external reservoir of particles which allows the 
total particle number number to fluctuate around some constant average 
value, then the total number of particles need only be conserved on the 
average.  Formally, one standard way to compensate for the lack of 
exact particle number conservation is to use the Lagrange multiplier 
method and subtract a chemical potential term $\mu N_{op}$ from the 
Hamiltonian (just as in statistical mechanics when one goes from the 
canonical ensemble to the grand canonical ensemble)~\cite{Hugenholtz}
\begin{equation}
H \rightarrow H'= H - \mu N_{op},
\end{equation}
where $N_{op}=\sum_{p}a_{p}^{\dagger}a_{p}$ is the total number 
operator, and $\mu$ represents the chemical potential, i.e., the 
average energy for adding a particle to the open system described by 
$H$.
In the 
present context, we are considering the case of a Fabry-Perot cavity 
with low, but finite, transmissivity mirrors which allow photons to 
enter and leave the cavity, due to an input light beam coming in from 
the left and an output beam leaving from the right.  This permits a 
realistic physical implementation of the external reservoir, since the 
Fabry-Perot cavity allows the total particle number inside the cavity to 
fluctuate due to particle exchange with the beams outside the cavity.  
However, the photons remain trapped inside the cavity long enough so 
that a thermalized condition is achieved after many photon-photon 
interactions (i.e., after many collisions), thus allowing the 
formation of a photon fluid.

It will be useful to separate out the zero-momentum components of the 
interaction Hamiltonian, since it will turn out that there is a 
macroscopic occupation of the zero-momentum state due to 
Bose condensation.  The prime on the sums
$\sum'_{p}$, $\sum'_{p\kappa}$, and $\sum'_{\kappa pq}$ in the
following equation denotes sums over momenta explicitly
excluding the zero-momentum state, i.e., all the running indices
$p$, $\kappa$, $q$,$p+\kappa$,$q-\kappa$ which are not explicitly
set equal to zero, are nonzero:  
\begin{eqnarray}
H_{int}&=&\frac{1}{2}V(0)a_{0}^{\dagger} 
a_{0}^{\dagger}a_{0}a_{0} + 
V(0){\sum_{p}}'a_{p}^{\dagger}a_{p}a_{0}^{\dagger}a_{0}+ \nonumber\\
&&{\sum_{p}}'\left(V(p)a_{p}^{\dagger}a_{0}^{\dagger}a_{p}a_{0} 
+\frac{1}{2}\left[V(p)a_{p}^{\dagger}a_{-p}^{\dagger}a_{0}a_{0} + 
V(p)a_{0}^{\dagger}a_{0}^{\dagger}a_{p}a_{-p}\right]\right)+\nonumber\\
&&{\sum_{p\kappa }}'V(\kappa ) \left(a_{p+\kappa 
}^{\dagger}a_{0}^{\dagger}a_{p}a_{\kappa } +a_{p+\kappa 
}^{\dagger}a_{-\kappa }^{\dagger}a_{p}a_{0}\right) 
+\frac{1}{2}{\sum_{\kappa pq}}'V(\kappa ) \left(a_{p+\kappa 
}^{\dagger} a_{q-\kappa }^{\dagger}a_{p}a_{q}\right)\,.
\label{10}
\end{eqnarray}
Here we have also assumed that $V(p)=V(-p)$.  By thus separating out 
the zero-momentum state from the sums in the Hamiltonian, and 
replacing all occurrences of $a_{0}$ and $a_{0}^{\dagger}$ by 
$\sqrt{N_{0}}$, we find that the Hamiltonian $H'$ decomposes into 
three parts
\begin{equation}
H'=H_{0} + H_{1} + H_{2}\,,
\end{equation}
where
\begin{equation}
H_{0} = \frac{1}{2}V(0)a_{0}^{\dagger} 
a_{0}^{\dagger}a_{0}a_{0} \approx \frac{1}{2}V(0)N_{0}^{2}\,, 
\label{H0}
\end{equation}
\begin{equation}
H_{1}\approx {\sum_{p}}'\epsilon'(p)a_{p}^{\dagger}a_{p} 
+\frac{1}{2}N_{0}{\sum_{p}}'V(p)\left(a_{-p}^{\dagger}a_{p}^{\dagger} 
+a_{-p}a_{p}\right)\,,
\label{original}
\end{equation}
\begin{equation}
H_{2}\approx \sqrt{N_{0}}\;{\sum_{p\kappa }}'V(\kappa  )
\left(a_{p+\kappa  }^{\dagger}a_{p}a_{\kappa  }
+a_{p+\kappa  }^{\dagger}a_{-\kappa  }^{\dagger}a_{p}\right)
+\frac{1}{2}{\sum_{\kappa  pq}}'V(\kappa  )
\left(a_{p+\kappa  }^{\dagger} a_{q-\kappa  }^{\dagger}a_{p}a_{q}\right)\,,
\end{equation}
where
\begin{equation}
\epsilon'(p) = \epsilon(p) + N_{0}V(0) + N_{0}V(p) - \mu
\label{primed}
\end{equation}
is a modified photon energy, and where $N_{0}$ and $\mu$ are given by
\begin{equation}
N_{0} + <\Psi_{0}| {\sum_{p}}'a_{p}^{\dagger}a_{p}
|\Psi_{0}> = N\,
\end{equation}
and
\begin{equation}
\mu =\frac{\partial E_{0}}{\partial N}\,.
\end{equation}
Here $E_{0}=\left<\Psi_{0}|H|\Psi_{0}\right>$ 
is the ground state energy of $H$.  
In the approximation that there is little depletion of the Bose 
condensate due to interactions (i.e., $N\approx N_{0}\gg1$), the first 
term of Eq.~(\ref{10}) (i.e., $H_{0}$ in Eq.~(\ref{H0})) dominates, so 
that
\begin{equation}
E_{0}\approx \frac{1}{2}N_{0}^{2}V(0)
\approx\frac{1}{2}N^{2}V(0),
\end{equation}
and therefore that
\begin{equation}
\mu\approx NV(0)\approx N_{0}V(0).
\label{chemical}
\end{equation}
This implies that the effective chemical potential of a photon, 
i.e., the energy for adding a photon to the 
photon fluid,
is given by the number of photons in the Bose condensate 
times the repulsive pairwise interaction energy between photons with 
zero relative momentum. 
It should be remarked that the fact that the chemical potential is 
nonvanishing here makes the thermodynamics of this two-dimensional 
photon system quite different from the usual three-dimensional, Planck 
blackbody photon system~\cite{blackbody}.  
In the same approximation, Eq.~(\ref{primed}) becomes
\begin{equation}
\epsilon'(p) \approx \epsilon(p) + N_{0}V(p).
\label{prime}
\end{equation}
This is the single-particle photon energy in the Hartree 
approximation.

In the same approximation, it is also assumed that 
$|H_{1}|\gg|H_{2}|$, i.e., that the interactions between the bosons 
are sufficiently weak, again so as not to deplete the Bose condensate 
significantly.  In the case of the weakly-interacting photon gas 
inside the Fabry-Perot resonator, since the interactions between the 
photons are indeed weak, this assumption is a good one.

Following Bogoliubov, we now introduce the following canonical 
transformation in order to diagonalize the quadratic-form 
Hamiltonian $H_{1}$ in Eq.~(\ref{original}):
\begin{eqnarray}
\alpha_{\kappa  }&=&u_{\kappa  }a_{\kappa  }+v_{\kappa  }a_{-\kappa  }^{\dagger}\nonumber \\
\alpha_{\kappa }^{\dagger}&=&u_{\kappa }a_{\kappa 
}^{\dagger}+v_{\kappa }a_{-\kappa  }\,.
\label{canonical}
\end{eqnarray}
Here $u_{\kappa  }$ and $v_{\kappa  }$ are two real $c$-numbers 
which must satisfy the condition
\begin{equation}
u_{\kappa  }^{2}-v_{\kappa  }^{2}=1\,,
\label{18}
\end{equation}
in order to insure that the Bose commutation relations are preserved 
for the new creation and annihilation operators for certain 
quasi-particles, $\alpha_{\kappa }^{\dagger}$ and $\alpha_{\kappa }$, 
i.e., that
\begin{equation}
[\alpha_{\kappa  },\alpha_{\kappa  '}^{\dagger}]=\delta_{\kappa  ,\kappa  '}
\mbox{ and }[\alpha_{\kappa  },\alpha_{\kappa  '}]= 
[\alpha_{\kappa  }^{\dagger},\alpha_{\kappa  '}^{\dagger}]=0\,.
\end{equation}
We seek a diagonal form of $H_{1}$ given by
\begin{equation}
H_{1}={\sum_{\kappa }}'\left[\tilde{\omega} (\kappa )\left(\alpha_{\kappa 
}^{\dagger}\alpha_{\kappa } 
+\frac{1}{2}\right)+\mbox{constant}\right]\,,
\label{diagonal}
\end{equation}
where $\tilde{\omega}(\kappa )$ represents the energy of a 
quasi-particle of momentum $\kappa $.  Substituting the new 
creation and annihilation operators $\alpha_{\kappa }^{\dagger}$ and 
$\alpha_{\kappa }$ given by Eq.~(\ref{canonical}) into 
Eq.~(\ref{diagonal}), and comparing with the original form of the 
Hamiltonian $H_{1}$ in Eq.~(\ref{original}), we arrive at the 
following necessary conditions for diagonalization:
\begin{eqnarray}
\tilde{\omega} (\kappa  )u_{\kappa  }v_{\kappa  }&=&\frac{1}{2}N_{0}V(\kappa  ) \label{21}\\
u_{\kappa  }^{2}&=& \frac{1}{2}\left[1+\epsilon'(\kappa  )/\tilde{\omega} (\kappa  )\right] \label{22}\\
v_{\kappa  }^{2}&=&\frac{1}{2}\left[-1+\epsilon'(\kappa  )/\tilde{\omega} (\kappa  )\right]. \label{23}
\end{eqnarray}
Squaring Eq.~(\ref{21}) and substituting from 
Eqs.~(\ref{22}) and (\ref{23}), we obtain
\begin{equation}
\tilde{\omega} (\kappa  )^{2}=\epsilon'(\kappa  )^{2}-N_{0}^{2}V(\kappa  )^{2}
= \epsilon(\kappa  )^{2}+ 2\epsilon(\kappa  )N_{0}V(\kappa  ),
\label{square}
\end{equation}
where in the last step we have used Eq.~(\ref{prime}).

\begin{figure}
\centerline{\psfig{figure=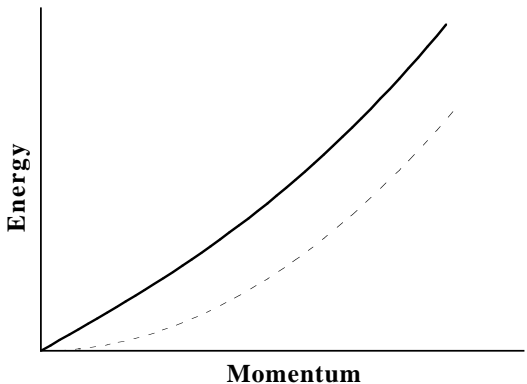,width=7cm}}
\caption{The energy versus momentum of an elementary excitation in the 
weakly-interacting Bose gas, here in the present case, the photon 
fluid.  The solid line represents the Bogoliubov dispersion relation 
given by Eq.~(\ref{28}), for the special case that $V(\kappa  ) = 
V(0) =$ constant, and the dashed line represents a quadratic 
dispersion relation for a noninteracting, diffracting photon inside 
the Fabry-Perot resonator.}
\label{Bogoliubov}
\end{figure}

Thus the final result is that the Hamiltonian $H_{1}$ in 
Eq.~(\ref{diagonal}) describes a collection of noninteracting simple 
harmonic oscillators, i.e., quasi-particles, or elementary 
excitations of the photon fluid from its ground state.  The 
energy-momentum relation of these quasi-particles is obtained from 
Eq.~(\ref{square}) upon substitution of $\epsilon(\kappa  )=\kappa  ^{2}/2m$ from 
Eq.~(\ref{7})
\begin{equation}
\tilde{\omega} (\kappa  ) = \left[\frac{\kappa  ^{2} N_{0} V(\kappa  )}{m} +
\frac{\kappa  ^{4}}{4m^{2}}\right]^{1/2},
\label{28}
\end{equation}
which we shall call the ``Bogoliubov dispersion relation.''  This 
dispersion relation is plotted in Fig.~\ref{Bogoliubov}, in the 
special case that $V(\kappa ) = V(0) =$ constant.  (Note that Landau's 
roton minimum could in principle also be incorporated into this theory 
by a suitable choice of the functional form of $V(\kappa )$.)

For small values of $\kappa $ this dispersion relation is {\em linear} 
in $\kappa $.  This feature, together with the fact that the operator 
$\alpha_{\kappa }^{\dagger}\alpha_{\kappa }$ in Eq.~(\ref{diagonal}) 
describes a density fluctuation in the fluid, indicates that the 
nature of the elementary excitations here is that of {\em phonons}, 
which in the classical limit of large phonon number leads to 
sound-like waves propagating inside the photon fluid at the sound 
speed
\begin{equation}
v_{s}= \lim_{\kappa \rightarrow 0}\frac{\tilde{\omega}(\kappa 
)}{\kappa} = 
\left(\frac{N_{0}V(0)}{m}\right)^{1/2}=\left(\frac{\mu}{m}\right)^{1/2}\,.
\label{29}
\end{equation}
At a transition momentum $\kappa  _{c}$ given by
\begin{equation}
\kappa  _{c}= 2\left(mN_{0}V(\kappa  _{c})\right)^{1/2}
\end{equation}
(i.e., when the two terms of Eq.~(\ref{28}) are 
equal), the linear relation between energy and momentum turns into a 
quadratic one, indicating that the quasi-particles at large momenta 
behave essentially like nonrelativistic free particles with an energy 
of $\kappa  ^{2}/2m$.  The reciprocal of $\kappa  _{c}$ defines a characteristic 
length scale
\begin{equation}
\lambda_{c}\equiv 2\pi\hbar /\kappa  _{c}=\pi\hbar /mv_{s}\,,
\end{equation}
which characterizes the distance scale over which collective effects 
arising from the pairwise interaction between the photons become 
important.

Thus in the above analysis, we have shown that all the approximations 
involved in the Bogoliubov theory should be valid ones for the case of 
the 2D photon fluid inside a nonlinear Fabry-Perot cavity.  Hence the 
Bogoliubov dispersion relation should indeed apply to this fluid; in 
particular, there should exist sound-like modes of propagation in the 
photon fluid.

\section{Classical Picture of Sound Waves in a Nonlinear Optical 
Fluid}

\begin{figure}
\centerline{\psfig{figure=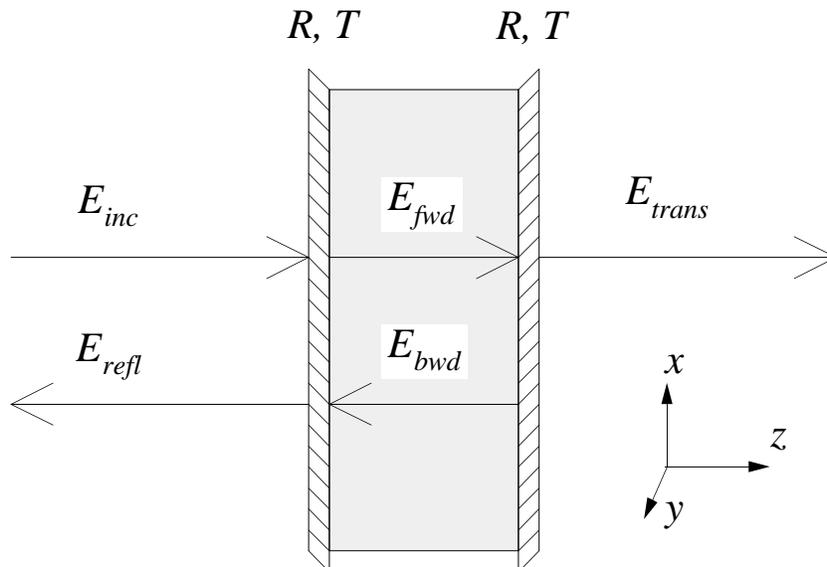,width=11cm}}
\caption{Fields and coordinate system in the Fabry-Perot cavity.  The 
applied field $E_{inc}$ arises from a laser beam incident from the 
left.  An atomic vapor excited to the red side of resonance by the 
incident light fills the space (the gray area) between the two 
mirrors.  The presence of these atoms leads to a self-defocussing Kerr 
nonlinearity (corresponding to repulsive photon-photon interactions) 
inside the cavity.}
\label{cavityfig}
\end{figure}

A classical nonlinear optical treatment of a Fabry-Perot cavity which 
is filled with a medium with a self-defocussing Kerr nonlinearity (see 
Fig.~\ref{cavityfig}), also indicates the existence of modes of 
sound-like wave propagation in the nonlinearly interacting light.  
Such a nonlinear medium could consist of an alkali atomic vapor 
excited by a laser detuned to the red side of resonance.  In fact, it 
turns out that fluctuations in the light intensity in this medium 
propagate with a dispersion relation which is identical to that given 
above in Eq.~(\ref{28}) for the weakly-interacting Bose gas.

To derive this dispersion relation classically, we begin by 
considering the planar Fabry-Perot cavity shown in 
Fig.~\ref{cavityfig}.  Two parallel planar mirrors of reflectivity $R$ 
and transmissivity $T$ (with $R+T=1$, i.e., with no dissipation) are 
normal to the $z$-axis and separated by a distance $L$.  A laser beam 
travelling in the $+z$ direction is incident on the cavity, and there 
results five interacting light beams in the problem.  The region 
between the mirrors (inside the cavity) contains a nonlinear 
polarizable medium.  The classical electric field obeys Maxwell's 
equations, written in wave-equation form in CGS units as
\begin{equation}
\frac{\partial^{2}{\mathrm E}}{\partial 
z^{2}}+\nabla_{\perp}^{2}{\mathrm E}
-\frac{1}{c^{2}}\frac{\partial^{2}{\mathrm E}}{\partial t^{2}} =
\frac{4\pi}{c^{2}}\frac{\partial^{2}{\mathrm P}}{\partial t^{2}}\,,
\label{02waveequation}
\end{equation}
where ${\mathrm E}$ is the (real) electric field amplitude, 
${\mathrm P}$ is the
polarization introduced in the medium, and $\nabla_{\perp}^{2}$ is
the Laplacian in the transverse coordinates $x$ and $y$.  This
equation is supplemented by boundary conditions at the two mirrors.

Equation~(\ref{02waveequation}) simplifies considerably when the
following assumptions are made:
\begin{enumerate}
\item The slowly-varying envelope approximation is justified, in
which case we recast Eq.~(\ref{02waveequation}) in terms of the
field envelope ${\cal E}$.
\item The frequency spacing between adjacent longitudinal cavity modes
is much greater than
\begin{enumerate}
	\item the incident laser linewidth, and
	\item the nonlinearity bandwidth,
\end{enumerate}
allowing us to neglect the $z$-dependence of the field envelope (this
is sometimes called the \textit{uniform field} approximation).
\item The atomic response time is much shorter than the cavity
lifetime, allowing us to adiabatically eliminate the atomic  response
(i.e., the nonlinearity is fast).
\end{enumerate}

Under these reasonable assumptions the cavity's internal field envelope 
is governed by the Lugiato-Lefever equation \cite{lugiato1987}, 
written here as
\begin{equation}
\frac{\partial \cal E}{\partial t} = 
\frac{i c}{2 k} \nabla_{\perp}^{2}{ \cal E}
+ i\omega n_2 |{ \cal E}|^2{ \cal E}
+ i(\Delta\omega){ \cal E}
- \Gamma({\cal E}- {\cal E}_{d})\,,
\label{dimcavityeqn}
\end{equation}
where ${\cal E}(x,y,t)$ is the internal cavity field envelope 
amplitude, $k$ is the longitudinal wavenumber, $\omega$ is the laser 
angular frequency, $n_2$ is the nonlinear index inside the cavity 
($n\approx 1+n_{2}|{\cal E}|^{2}$), $\Delta\omega=\omega-\omega_{cav}$ 
is the detuning of the driving laser from linear cavity 
resonance, $\Gamma=cT/2L$ is the cavity decay rate, and ${\cal 
E}_{d}(x,y)$ is a driving laser amplitude.  In other contexts, 
Eq.~(\ref{dimcavityeqn}) is called the Nonlinear Schr\"{o}dinger (NLS) 
equation, or the Ginzburg-Landau equation, or the Gross-Pitaevskii 
equation.  The latter two of these were introduced as descriptions of 
superfluid and of Bose-Einstein-condensed systems, with a complex order 
parameter $\Psi$, which here is identified with $\cal E$.

Equation~(\ref{dimcavityeqn}) has the nonlinear plane-wave solution
 \begin{equation}
{\cal E} = {\cal E}_{0}\,\exp[i (\omega n_2 { \cal E}_{0}^2+ 
\Delta\omega)t]
\label{nonlinear.eigenmode}
\end{equation}
when $\Gamma$ is negligible~\cite{Perez-Torres}, 
in which case ${\cal E}_{0}$ can be assumed real without loss of 
generality.
Linearizing around this solution by substituting 
the form
\begin{equation}
{\cal E} = \left({\cal E}_{0}+a(x,y,t)\right)\,
\exp[i (\omega n_2 { \cal E}_{0}^2+
\Delta\omega)t],
\end{equation}
we get the following linear equation for the fluctuation amplitude
(we have assumed that $|a(x,y,t)|\ll {\cal E}_{0}$):
\begin{equation}
\frac{\partial a}{\partial t} = \frac{i c}{2 k} \nabla_{\perp}^{2}a +
i\omega n_{2}{\cal E}_{0}^{2} (a + a^{*})\,.
\label{linearizedeqn}
\end{equation}

\begin{figure}
\centerline{\psfig{figure=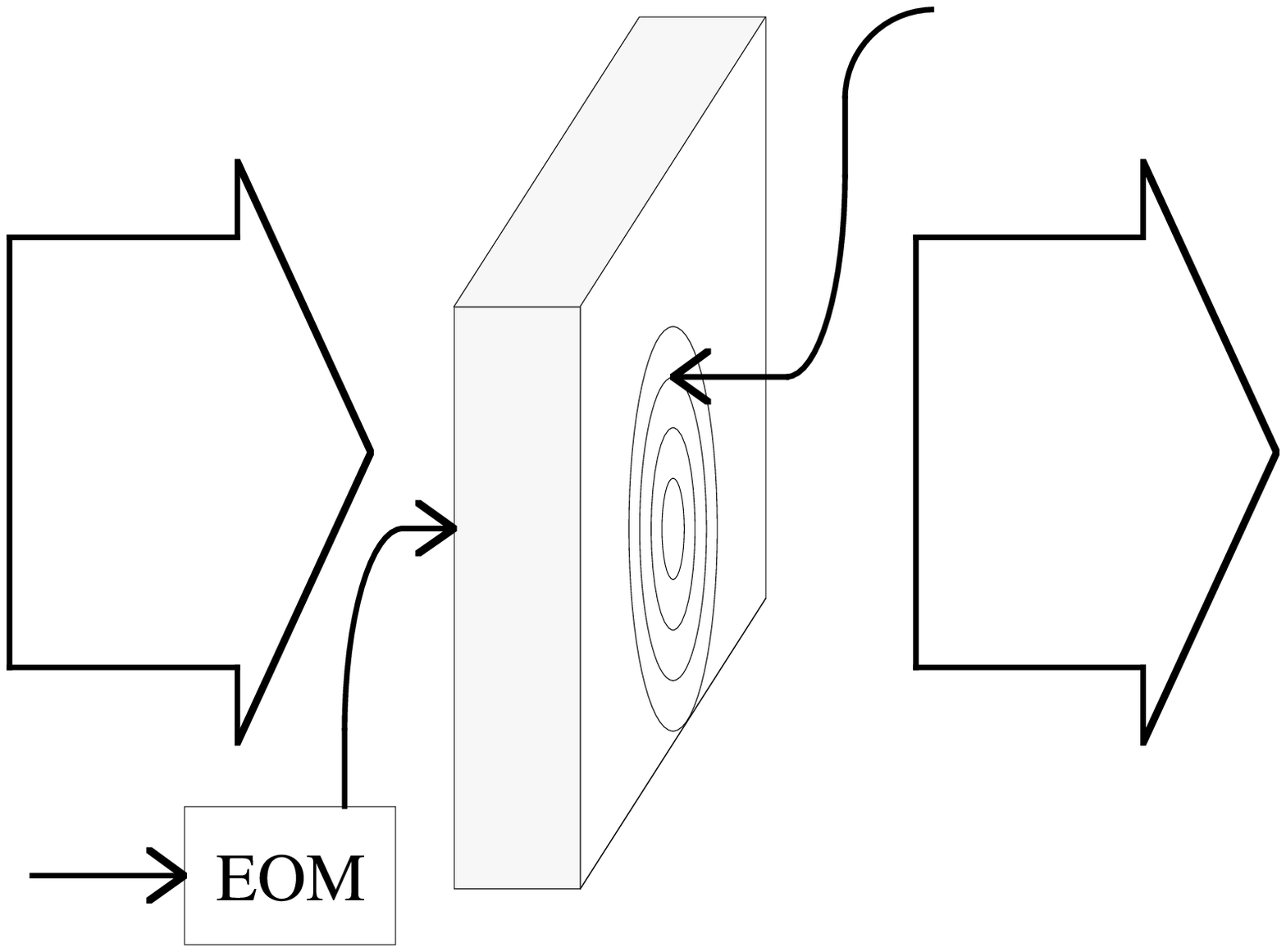,width=7cm}}
\caption{Schematic of an experiment to observe the sound-like waves in a 
photon fluid which fills a nonlinear Fabry-Perot resonator.  The 
nonlinear medium (denoted by the gray area) is an alkali atomic vapor 
excited by a broad laser beam (denoted by the broad incoming arrow) 
tuned to the red side of resonance.  The goal is the verify the 
Bogoliubov dispersion relation, Eq.~(\ref{28}) or (\ref{Jack}).  An 
electro-optic modulator (EOM) modulates the intensity of light at a 
radio frequency in the MHz range, which is then injected by means of 
an optical fiber tip at a single point on the entrance face of the 
Fabry-Perot resonator.  The wavelength of the resulting sound-like 
waves can be measured by scanning in the transverse direction the tip 
of another optical fiber across the output face of the Fabry-Perot, 
and by measuring the phase of the modulated pick-up signal relative to 
that of the EOM modulation signal.}
\label{experimentfigure}
\end{figure}

Here we look for a cylindrically symmetric solution appropriate for 
the experimental geometry (see Fig.~\ref{experimentfigure}).  
Substituting the trial solution
\begin{equation}
a(\rho,t)=\alpha J_{0}(K\rho)e^{i\Omega^{*} t}+
\beta J_{0}(K\rho)e^{-i\Omega t},
\end{equation}
where $J_{0}(K\rho)$ is the zero-order Bessel function, $\rho 
=(x^{2}+y^{2})^{1/2}$ is the transverse radial distance from the 
origin of a fluctuation, and $K$ is the wavenumber of the fluctuation, 
we obtain the following dispersion relation for small-amplitude 
intensity fluctuations in the light filling the cavity~\cite{CKG}:
\begin{equation}
	\Omega(K)=\left[c^{2}K^{2}\left|n_{2}\right|{\cal E}_{0}^{2}+
	\frac{c^{4}K^{4}}{4\omega^{2}}\right]^{1/2}\,,
\label{Jack}
\end{equation}
where $\Omega$ and $K$ are the angular frequency and wavenumber 
respectively of the transverse sound-like mode. 

For transverse wavelengths much longer than $\Lambda_{c}\equiv 
\lambda/\left(\Delta n\right)^{1/2}$, where $\lambda$ is the optical 
wavelength and $\Delta n=\left|n_{2}\right|{\cal E}_{0}^{2}$ is the 
nonlinear index shift induced by the background beam, the transverse 
mode propagates with the constant phase velocity

\begin{equation}
v_{s} = c\sqrt{\Delta n}=c\sqrt{\left|n_{2}\right|{\cal E}_{0}^{2}}\;,	
\end{equation}

\noindent which we identify as a sound-wave velocity.  
This velocity 
is identical to the one found earlier in Eq.~(\ref{29}) for the 
velocity of phonons in the photon fluid, provided that one identifies 
the energy density of the light inside the cavity with the number of 
photons in the Bose condensate as follows:
\begin{equation}
{\cal E}_{0}^{2} = 8\pi N_{0}\hbar \omega/{V}_{cav}\;,
\label{40}
\end{equation}
where ${V}_{cav}$, the cavity volume, is also the 
quantization volume for the electromagnetic field, and provided that 
one makes use of the known proportionality between $n_{2}$ and 
$V(0)$~\cite{Deutsch,Akhmanov}.

In fact, the entire dispersion relation, Eq.~(\ref{Jack}), found above 
classically for sound-like waves associated with fluctuations in the 
light intensity inside a resonator filled with a self-defocusing Kerr 
medium, is formally identical to the Bogoliubov dispersion relation, 
Eq.~(\ref{28}), obtained quantum mechanically for the elementary 
excitations of the photon fluid, in the approximation $V(\kappa ) = 
V(0) =$ constant.  This is a valid approximation, since the pairwise 
interaction potential between two photons is given by a transverse 2D 
pairwise spatial Dirac delta function, whose strength is proportional 
to $n_{2}$~\cite{Deutsch,Akhmanov}.  It should be kept in mind that 
the phenomena of self-focusing and self-defocusing in nonlinear optics 
can be viewed as arising from \textit{pairwise interactions} between 
photons when the light propagation is paraxial and the Kerr 
nonlinearity is fast~\cite{Deutsch,Akhmanov}.  Since in a quantum 
description the light inside the resonator is composed of photons, and 
since these photons as the constituent particles are weakly 
interacting repulsively with each other through the self-defocusing 
Kerr nonlinearity to form a photon fluid, this formal identification 
is a natural one.

\section{An Experiment in Progress}

We are in the process of investigating experimentally the existence of 
the sound-like propagating photon density waves predicted above for a 
planar Fabry-Perot cavity containing a self-defocusing ($n_{2}<0$) 
nonlinear medium (see Fig.~\ref{experimentfigure}).

The sound-like mode is most simply observed by applying two incident 
optical fields to the nonlinear cavity: a broad plane wave resonant 
with the cavity to form the nonlinear background fluid on top of which 
the sound-like mode can propagate, and a weaker amplitude-modulated 
beam which is modulated at the sound wave frequency in the radio range 
by an electro-optic modulator, and injected by means of an optical 
fiber tip at a single point on the entrance face of the Fabry-Perot.  
The resulting weak time-varying perturbations in the background light 
induce transversely propagating waves in the photon fluid, which 
propagate away from the point of injection like ripples on a pond.  
This sound-like mode can be phase-sensitively detected by another 
fiber tip placed at the exit face of the Fabry-Perot some transverse 
distance away from the injection point, and its sound-like wavelength 
can be measured by scanning this fiber tip transversely across the 
exit face.

The experiment employs a cavity length $L$ of $2$ cm and mirrors with 
intensity reflectivities of $R=0.997$, for a cavity finesse of roughly 
$1000$.  The optical nonlinearity is provided by rubidium vapor at 
$80^{\rm o}$ C, corresponding to a number density of $10^{12}$ 
rubidium atoms per cubic centimeter.  We use a circularly-polarized 
laser beam, detuned by around $600$ MHz to the red side of the 
$^{87}\rm Rb$, $F=2\,\rightarrow\,F'=3$ transition of the $D_{2}$ 
line; using this closed transition eliminates optical pumping into the 
$F=1$ ground
state.  
This 600 MHz detuning of the laser from the atomic resonance is 
considerably larger than the Doppler width of 340 MHz, and the 
residual absorption arising from the tails of the nearby resonance 
line gives rise to a loss which is less than or comparable to the loss
arising from the mirror transmissions.  
This extra absorption loss contributes to a slightly larger effective 
cavity loss coefficient $\Gamma$, but does not otherwise alter the 
qualitative behavior of the Bogoliubov dispersion relation, nor any 
of the other main conclusions of this paper.  
The above criteria (1-3) for the validity of 
Eq.~(\ref{dimcavityeqn}), as well as those for the validity of the 
microscopic Bogoliubov theory, should be well satisfied by these 
experimental parameters.  An intracavity intensity of 
$40\,\mathrm{W/cm^{2}}$ results in $\Delta n=2\times10^{-6}$, for a 
sound speed $v_{s}=4.2\times10^{7}\,\mathrm{cm/s}$ and transition 
wavelength $\Lambda_{c}\approx 1\,\mathrm{mm}$.  For this intensity, 
$N_{0} \approx 8\times 10^{11}$, so that the condition for the 
validity of the Bogoliubov theory, $N_{0}\gg1$, is well satisfied.

\section{Discussion and Future Directions}

We suggest here that the Bogoliubov form of dispersion relation, 
Eq.~(\ref{28}) or (\ref{Jack}), implies that the photon fluid formed 
by repulsive photon-photon interactions in the nonlinear cavity is 
actually a photon {\em superfluid}.  This means that a superfluid 
state of light might actually exist.  
Although the exact definition of superfluidity is 
presently still under discussion, especially in light of the question 
whether the recently discovered atomic Bose-Einstein condensates 
are superfluids or not~\cite{Walls}, one indication of the 
existence of a photon superfluid would be that there exists 
a 
critical transition from a 
dissipationless state of superflow, i.e., a laminar flow of the photon 
fluid below a certain critical velocity past an obstacle, into a 
turbulent state of flow, accompanied by energy dissipation associated 
with the shedding of von-Karman-like $quantized$ vortices past 
this obstacle, above this critical velocity.
(It is the generation of {\em quantized} vortices above this critical 
velocity which distinguishes the onset of {\em superfluid} turbulence 
from the onset of {\em normal} hydrodynamic turbulence.)

The Bogoliubov dispersion relation (plotted earlier in 
Fig.~\ref{Bogoliubov}) consists of two regimes: (1)~a linear regime, 
in which there is a linear relationship between the energy of the 
elementary excitation and its momentum near the origin (i.e., for low 
energy excitations) corresponding to the sound-like waves, or more 
precisely, to the phonons in the photon fluid, produced by the 
collective oscillations of this fluid, in which the photons are 
coupled to each other by the mutually repulsive interactions between 
them, and (2) a quadratic regime, in which there is a quadratic 
relation for sufficiently large transverse momenta corresponding to 
the diffraction of the component photons, which would dominate when 
the pairwise interactions between the photons can be neglected.  A 
crude one-dimensional model can give rise to an understanding of the 
origin of the sound-like waves in the photon fluid: Consider a system 
consisting of identical steel balls placed on a frictionless track.  
This system of balls is initially motionless.  Now set a ball at the 
one end of the track into motion so that it collides with its nearest 
neighbor.  The momentum transfer between adjacent hard spheres on this 
track, as they collide with one another, sets up a moving pattern of 
density fluctuations among the balls, which propagates like a sound 
wave from one end of the track towards the other end.  Such a 
sound-like wave carries energy and momentum with it as it propagates.

It may be asked why the classical nonlinear optical calculation gives 
the same result as the quantum many-body calculation.  One 
answer is that one expects classical sound waves to have the same 
dispersion relation as phonons in a quantum many-body system: there 
exists a classical, correspondence-principle limit of the quantum 
many-body problem, in which the collective excitations (i.e., their 
dispersion relation) do not change their form in the classical limit 
of large phonon number.

The physical meaning of this dispersion relation is that the lowest 
energy excitations of the system consist of quantized sound waves or 
phonon excitations in a superfluid, whose maximum critical velocity is 
then given by the sound wave velocity.  By inspection of this 
dispersion relation, a single quantum of any elementary excitation 
cannot exist with a velocity below that of the sound wave.  Hence no 
excitation of the superfluid at zero temperature is possible at all 
for any object moving with a velocity slower than that of the sound 
wave velocity, according to an argument by Landau~\cite 
{Landau}.  Hence the 
flow of the superfluid must be dissipationless below this critical 
velocity.  Above a certain critical velocity, 
dissipation due to vortex shedding is expected from 
computer simulations based on the Gross-Pitaevskii (or Ginzburg-Landau 
or nonlinear Schr\"{o}dinger) equation which should give an accurate 
description of this system at the macroscopic level~\cite{Pomeau}.

We propose a follow-up experiment to demonstrate that the sound wave 
velocity, typically a few thousandths of the vacuum speed of light, 
is indeed a maximum critical velocity of a fluid, i.e., that 
this photon fluid exhibits persistent currents in accordance with the 
Landau argument based on the Bogoliubov dispersion relation.  
Suppose we shine light at some nonvanishing incidence angle on a 
Fabry-Perot resonator (i.e., exciting it on some off-axis mode).  This 
light produces a uniform flow field of the photon fluid, which flows 
inside the resonator in some transverse direction and at a speed 
determined by the incidence angle.  A cylindrical obstacle placed 
inside the resonator will induce a laminar flow of the superfluid 
around the cylinder, 
as long as the flow velocity remains 
below a certain critical velocity.  
However, above this critical velocity a 
turbulent flow will be induced, with the formation of a von-Karman 
vortex street associated with quantized vortices shed from the 
boundary of the cylinder~\cite{Pomeau}.  The typical vortex core size 
is given by the light wavelength divided by the square root of the 
nonlinear index change.  Typically the vortex core size should 
therefore be around a few hundred microns, so that this nonlinear 
optical phenomenon should be readily observable.

\section*{Acknowledgments} We thank L.M.A. Bettencourt, D.A.R. Dalvit, 
I.H. Deutsch, J.C. Garrison, D.H. Lee, M.W. Mitchell, J. Perez-Torres, 
D.S. Rokhsar, D.J. Thouless, E.M. Wright, and W.H. Zurek for helpful 
discussions.  The work was supported by the ONR and by the NSF.

\pagebreak

\section*{FIGURES}

\begin{thebibliography}{99}
\bibitem{blackbody}
It may be asked how this photon gas problem in two dimensions is 
different from the Planck blackbody problem in three dimensions.  The 
first part of the answer is that the Fabry-Perot resonator makes the 
problem effectively two-dimensional, since the resonator is excited by 
a nearly monochromatic laser beam with a narrow linewidth, which 
selects out only a single longitudinal mode of the Fabry-Perot to be 
excited.  Thus the $z$ degree of freedom for the photons inside the 
resonator is eliminated from the problem.  The second part of the 
answer is that in contrast to the Planck problem, where the chemical 
potential of the photon vanishes, here there is a nonvanishing 
chemical potential, Eq.~(\ref{chemical}), of the photon in the photon 
fluid, which arises from the repulsive pairwise interactions between 
photons in the Bose condensate inside the Fabry-Perot resonator.  For 
the same reasons, a two-dimensional phase transition of the 
Kosterlitz-Thouless type should be possible in the photon fluid.
\end{thebibliography}
\end{document}